\documentstyle[prd,aps,eqsecnum,psfig]{revtex}
\def\evat{\mathop{\big|}\limits}
\newcommand{\be}{\begin{equation}}
\newcommand{\ee}{\end{equation}}
\newcommand{\ben}{\begin{equation*}}
\newcommand{\een}{\end{equation*}}
\newcommand{\bea}{\begin{eqnarray}}
\newcommand{\eea}{\end{eqnarray}}
\newcommand{\beax}{\begin{eqnarray*}}
\newcommand{\eeax}{\end{eqnarray*}}

\newcommand{\kbm}{\begin{mathletters}}
\newcommand{\kem}{\end{mathletters}}

\begin{document}
\draft
\title{Cross Section and Effective Potential in 
Asymptotically Free Scalar Field Theories}
\author{Kenneth Halpern\thanks{This work was supported in part by 
funds provided by the U.S. Department 
of Energy under cooperative agreement \# DE-FC02-94ER40818.  The author
thanks Kerson Huang for his guidance in this research.}}
\address{}
\date{\today}
\maketitle
\begin{abstract}
In an effort to understand the physical implications of the newly discovered
non-trivial directions in scalar field theory, we compute lowest order
scattering amplitudes, cross sections, and the $1$-loop 
effective potential.  To lowest order, 
the primary effect of the nonpolynomial nature of the theories
is a renormalization of the field.  The high energy scaling of the 
$2\rightarrow 2$ cross section is studied and found to differ 
significantly from that of pure
$\phi^4$ theory.  From the $1$-loop effective potential we determine that in 
some cases radiative corrections destroy classical symmetry breaking, resulting
in a phase boundary between symmetry broken and unbroken theories.
No radiatively induced symmetry breaking is observed.

\end{abstract}
\section{Introduction}

In previous papers\cite{huang1,huang2}, we demonstrated the existence of 
asymptotically free approaches to the Gaussian fixed point of
scalar field theory, corresponding to nonpolynomial potentials. 
This calculation has been verified by 
Periwal\cite{princeton} using Polchinski's formulation\cite{polchinski} 
of the renormalization group.  We studied the space of all actions 
consisting of local, non-derivative interactions $S= \int d^dx [{1\over 2}
(\partial\phi)^2+U(\phi^2)]$ with $U(\phi^2)= \sum_{n=1}^{\infty} u_{2n} 
\phi^{2n}$.  Using the Wegner-Houghton\cite{wegner} renormalization 
group equations, we discovered a previously unexamined class of 
eigendirections to the Gaussian fixed point.  The potential associated with 
an eigenvalue $\lambda$ (labeled for convenience by $a\equiv {\lambda-2\over
d-2}$) is 
$$U_a(\phi^2(x))= r{N S_d\over 2(a-1) (d-2)}
\left[M(a-1, N/2, {(d-2)\phi^2\over 2S_d})-1\right]$$ 
where $M(a,b,z)$ is Kummer's function\footnote{%
$M(a,b,z)= {(b-1)!\over (a-1)!}\sum_{n=0}^{\infty} {z^n\over n!}
{(a+n-1)!\over (b+n-1)!}$.}(a type of 
confluent hypergeometric function), $S_d= {2^{1-d}\pi^{-{d\over 2}}
\over\Gamma({d\over 2})}$, $N$ is the number of field components, $d$ is the 
space-time dimension, and $r$ is the parameter representing
``distance'' along the eigendirection\footnote{Recall $r=2u_2$.}.  
The coefficients $u_{2n}$ are
given by $u_{2n}= {r\over 2}\Big({d-2\over 2S_d}\Big)^{n-1}
{\Gamma(a+n-1)\Gamma({N\over 2}+1)\over \Gamma(a)\Gamma(n+{N\over 2})n!}$.
The theory scales along an eigendirection as $r(t)= r_0 e^{\lambda t}$, with 
$t<0$ the direction of increasing cutoff.  Asymptotically free theories
correspond to $\lambda>0$ ($a>-1$ when $d=4$).  In $d=4$, for $-1\leq a<0$ 
these theories also exhibit symmetry breaking on the classical level.   

In this paper, we examine the lowest order scattering cross section and 
the $1$-loop effective potential for a single component scalar field theory 
in $4$ space-time dimensions\footnote{The extension to 
arbitrary components and
dimension is straightforward.  For the case $N=1$, $d=4$:  
$u_{2n}= r\big({4\over S_4}\big)^{n-1}
{\Gamma(a+n-1)\over (2n)!\Gamma(a)}$ and $S_4= {1\over 8 \pi^2}$.}.  
These are among the few tractable
physical calculations using our nonpolynomial potentials.  They
represent a first step towards understanding the implications of our 
eigendirections for the Higgs sector of the Standard Model.  
Perturbation theory and Feynman diagrams are designed for theories 
with a finite number of 
vertices.  Ordinary perturbative calculations in a nonpolynomial 
theory are intractable beyond lowest order.  An
entirely new perturbative machinary is necessary to perform
higher order computations. 
The $1$-loop effective potential is calculable using 
Jackiw's functional approach\cite{jackiw}.  In such functional methods lies
our greatest hope for analyzing nonpolynomial theories.  
We work in Euclidean space-time and employ
dimensionless parameters, fields, and momenta (i.e we set the cutoff
$\Lambda=1$).    

\section{Scattering Amplitudes: Unbroken Theories ($\char97>0$ , $\char114>0$)}

\begin{figure}
\centerline{\hbox{\hskip 0.2truein
\psfig{file=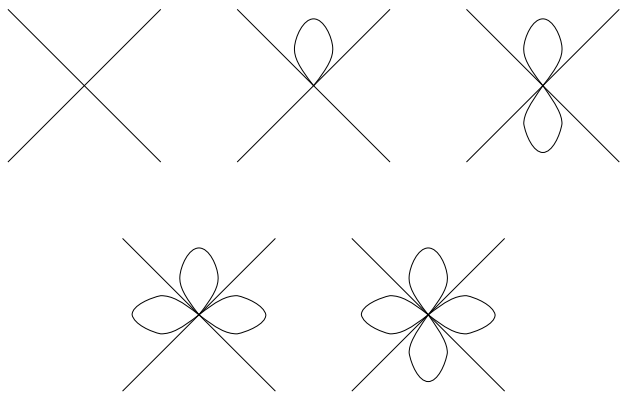,height=2truein,angle=-90}}}
\caption{Contributors to the 4-point amplitude.}
\label{fig:diagrams}
\end{figure}

The lowest order calculation of scattering amplitudes
involves a sum of diagrams as shown in 
figure (\ref{fig:diagrams}).  To an amplitude with $2n$ lines, 
there are contributions from vertices with $2m\geq 2n$ lines and $m-n$ 
internal contractions. 

For a $2n$-point amplitude , there are ${(2m)!\over (2n)!(2m-2n)!}$ ways
to choose which lines remain uncontracted, ${(2m-2n)!\over 2^{m-n}(m-n)!}$ 
ways to contract the remaining lines amongst themselves, and $(2n)!$ ways
to connect to the external lines.  The sum we need compute is ($n>1$)

\be
\label{eq:newvertex}
A_{2n}= \sum_{m=n}^{\infty} u_{2m} I^{m-n}{(2m)!\over 2^{m-n}(m-n)!}
\ee 

\noindent where $I$ is the internal contraction.  $I$ is independent of 
the external momenta and may be computed

\be
\label{eq:Idef}
I\equiv\int_{0}^{1} {d^4k\over (2\pi)^4} {1\over k^2+r}= 
{S_4\over 2} \Big(1+r\ln{r\over 1+r}\Big)
\ee

\noindent We substitute $u_{2n}$ from our eigenpotential and sum the series 
\ref{eq:newvertex} to obtain\footnote{This result requires that 
$|1+r\ln{r\over 1+r}|<1$, which is satisfied for the range under
consideration $1>r>0$.} 

\be
\label{eq:n1newvert}
A_{2n}= r 2^{2n-2} (S_4)^{1-n}{\Gamma(a+n-1)\over \Gamma(a)}
\Big(1-{2I\over S_4}\Big)^{1-a-n}
\ee

\noindent which may be written more succinctly as

\be
\label{eq:u2nprimeu2n}
A_{2n}= u_{2n}(2n)!\Big(-r\ln {r\over 1+r}\Big)^{1-a-n}
\ee

The primary effect of 
lowest order summation is a scaling of the field.  Defining 

$$\phi_r\equiv {\phi\over \sqrt{\Big(-r\ln {r\over 1+r}\Big)}}, $$

\noindent the renormalized amplitudes are

\be 
\label{eq:u2nprimeu2nrenorm}
A^r_{2n}= u_{2n}(2n)!\Big(-r\ln {r\over 1+r}\Big)^{1-a}
\ee

\noindent As $r\rightarrow 0$ the renormalized amplitudes vanish, 
demonstrating the expected asymptotic freedom.

\section{Broken Potentials}

For $-1\leq a<0$, 
the eigenpotentials exhibit symmetry breaking at the classical
level. For these, we must    
compute the broken potential, expanded around one of the classical minima,
and use this as the basis for scattering calculations.  
Let $\rho$ be the location
of the minimum, defined by 
${dU\over d\phi}|_\rho= 0$ (this depends on $a$ but not on $r$),  
and let $\phi'$ be our new dynamical field.
$$U(\phi)= U(\rho+\phi')\equiv V(\phi')$$
$$V(\phi')= \sum_{n=0}^{\infty} v_n \phi'^n$$
\be
\label{eq:vprime}
v_n=  {1\over n!}{d^n U(\phi'+\rho)\over d\rho^n}\evat_{\phi'=0}= 
{1 \over n!}\sum_{m=[{n+1\over 2}]}^{\infty} u_{2m}{(2m)!\over (2m-n)!}
\rho^{2m-n}
\ee

\noindent where $[x]$ is the greatest integer $\leq x$.  
Because odd vertices arise, we must examine 
$v_{2n}$ and $v_{2n+1}$ separately.  
These may be obtained either by summing the series or by employing the 
differential properties of Kummer functions (\cite{abramowitz}, 13.4.9):

\kbm
\label{eq:vnformulas}

\be
v_{2n}= u_{2n}M\big(a+n-1,{1\over 2},{\rho^2\over S_4}\big)
\ee

\be
v_{2n+1}= \rho u_{2n}{4 (a+n-1)\over S_4 (2n+1)}M\big(a+n,
{3\over 2},{\rho^2\over S_4}\big)
\ee

\kem

For convenience, we define

\be
\bar r\equiv 2 v_2= r M\big(a,{1\over 2},{\rho^2\over S_4}\big)
\ee

\section{Scattering Amplitudes: Broken Theories 
($-1\leq\char97<0$ , $\char114<0$)}

To compute the scattering amplitude for symmetry broken theories, 
we follow the same procedure
as before but use the $v_n$ as our vertices,  
instead of the $u_{2n}$.  The pertinent sum of diagrams is

\be
A_{n}= \sum_{j=0}^{\infty} v_{n+2j}I'^j{(n+2j)!\over 2^j j!}
\label{eq:vp2n}
\ee

with

\be
I'\equiv\int_{0}^{1} {d^4k\over (2\pi)^4} {1\over k^2+\bar r}= 
{S_4\over 2} \Big(1+\bar r\ln{\bar r\over 1+\bar r}\Big)
\ee

Performing the diagrammatic sum we obtain

\kbm
\label{eq:vprimefinal}

\be
A_{2n}= r {S_4^{1-n}2^{2n-2}(a+n-2)!\over (a-1)!}
\Big(-\bar r \ln {\bar r\over 1+\bar r}\Big)^{1-a-n}
M\big(a+n-1,{1\over 2},
{\rho^2\over -S_4 \bar r\ln{\bar r\over 1+\bar r}}\big) 
\ee

\be
A_{2n+1}= r \rho{S_4^{-n}2^{2n}(a+n-1)!\over(a-1)!}
\Big(-\bar r \ln {\bar r\over 1+\bar r}\Big)^{-a-n}M\big(a+n,{3\over 2},
{\rho^2\over -S_4 \bar r\ln{\bar r\over 1+\bar r}}\big)
\ee

\kem

As in the unbroken case, the primary effect of summation is a renormalization 
of the field.  Defining a renormalized wave function
$$\phi_r= {\phi\over\sqrt{\big(-\bar r \ln {\bar r\over 1+\bar r}\big)}},$$ 
\noindent we obtain (denoting by $v_n(\rho)$ the coefficients' functional
dependence on the location of the minimum)\footnote{we 
use the unrenormalized $\rho$ in $\bar r$ 
because corrections are higher order.}

\be 
A^r_{n}= v_n(\rho_r)n!\Big(-\bar r\ln{\bar r\over 1+\bar r}\Big)^{1-a}
\ee

Note that $\rho_r$ rather than $\rho$ 
appears in $v_n$.  As $r\rightarrow 0$, $\rho_r\rightarrow\infty$.  This 
remaining divergence has a physical explanation.  When a polynomial 
theory is broken, it is possible to isolate the physics near the different
minima.  For small enough excitations, the wells are blind to one another. 
Our potentials are exponential for large $\phi$, so the walls are steep.
Scattering amplitudes calculated near one of the minima see contributions
from $v_n$ with arbitrarily large $n$.  Because $v_n$ is the $n^{th}$ 
derivative of the potential, this means that the scattering amplitude 
involves derivatives of all orders and is not ``local'' in field space.  
Regardless of how large a hump separates the wells and how small our 
excitations are, the wells can see one another.  We cannot construct a 
symmetry broken theory in which the physics arising from the two vacua
are isolated.  
	
As $r\rightarrow 0$, the depths of the minima vanish as $O(r)$ while the 
width of the hump separating them remains fixed.  Consequently,
tunneling effects become progressively more significant.  The inability of 
perturbation theory to account for these tunneling effects manifests itself as
an exponential divergence in the broken theory's scattering amplitudes.

\section{Scattering Cross Sections}

We are now in a position to compute scattering 
cross sections\footnote{Our theory has one particle type, so decays are 
kinematically disallowed.}.  At the order to which we have calculated, the 
scattering amplitudes contain no momentum dependence, and the cross sections 
are purely kinematic.
As an example, we determine the cross section for $2\rightarrow 2$ scattering. 
Given a total energy of 
$E$, and working in the center of mass frame,

\be 
\label{eq:scattering22}
\sigma_{2\rightarrow 2}(E)= {1\over 2}\Big({1\over 2 E 
\sqrt{E^2-4 m^2}}\Big)
(A^{r}_4)^2
\int_{0}^{1}{d^3p_1\over (2\pi)^3 2 \omega_{p_1}}
{d^3p_2\over (2\pi)^3 2 \omega_{p_2}}(2\pi)^4\delta^{3}(\vec p_1+\vec p_2)
\delta(\omega_{p_1}+\omega_{p_2}-E)
\ee

Performing the integrations and 
substituting our unbroken scattering amplitudes

\be
\label{eq:sigmaue}
\sigma_{2\rightarrow 2}(E)= {32 a^2 r^2 \pi^3\over E^2}
\Big(-r\ln {r\over 1+r}\Big)^{2-2a}   
\ee

The result is similar for the broken case

\be
\label{eq:sigmave}
\sigma_{2\rightarrow 2}(E)= {32 a^2 r^2 \pi^3\over E^2}
M(a+1,{1\over 2},{\rho_r^2\over S_4})^2
\Big(-\bar r\ln {\bar r\over 1+\bar r}\Big)^{2-2a}   
\ee

Our potential corresponds to an eigenapproach to the Gaussian fixed point. 
We choose a point $r_0$ 
along the trajectory with associated physical energy $E_0$ as our reference.
This choice is arbitrary;
we cannot fix it within our theory.  The mass scale changes under the RG 
procedure, so the energy scales as $E=E_0e^t$.  The scaling of 
$r$ with energy $E$ is given by\footnote{$\bar r$ scales the same way.} 

$$r(E)= r_0\Big({E\over E_0}\Big)^{\lambda}$$

For $r$ small ($E$ large), 
we may approximate $r\ln {r\over 1+r}\approx r\ln r$ and the scaling 
behavior of the cross section for large $E$ is

\be
\sigma_{2\rightarrow 2}(E)\approx ({8 (\lambda-2)^2 \pi^3}
r_0^{6-\lambda}E_0^{4-\lambda}\lambda^{4-\lambda})E^{\lambda^2-6\lambda-2}
(\ln E)^{4-\lambda}
\ee

and for the broken case

\be
\sigma_{2\rightarrow 2}(E)\approx ({8 (\lambda-2)^2 \pi^3}
r_0^{6-\lambda}E_0^{4-\lambda}\lambda^{4-\lambda})
M(a+1,{1\over 2},{\rho^2\over S_4})^2
M(a,{1\over 2},{\rho^2\over S_4})^{4-\lambda}
E^{\lambda^2-6\lambda-2}
(\ln E)^{4-\lambda}
\ee

which may be summarized for both cases by (with $c$ some constant)

\be 
\sigma_{2\rightarrow 2}(E)\sim c E^{\lambda^2-6\lambda-2}
(\ln E)^{4-\lambda}
\ee

This result only makes sense for 
$6.32>\lambda>-.317$.  Outside this range, the cross section diverges, and 
our calculation is invalid.  

\section{$1$-loop effective potential}

To compute the $1$-loop effective potential we use
Jackiw's functional method \cite{jackiw} (see eg. \cite{huang3}, 10.4).  
With $\Lambda=1$, 
the $1$-loop effective potential is 

\be
V(\phi)= U(\phi)+ {1\over 32\pi^2}U''(\phi)+ {U''(\phi)^2\over 
64\pi^2}\Big(-{1\over 2}+\ln U''(\phi)\Big)
\ee

The first two terms are $O(r)$ while the last term has pieces
that are $O(r^2\ln |r|)$ and $O(r^2)$.  We can render the $r$ dependence
explicit by defining

\be 
V(\phi)\equiv r h_0(\phi)+ (r^2\ln |r|) h_1(\phi)+ r^2 h_2(\phi)
\ee

\be
U(\phi)\equiv r f(\phi)
\ee

\noindent with

$$h_0= f+{1\over 32\pi^2}f''$$
$$h_1= {(f'')^2\over 64\pi^2}$$
$$h_2= {(f'')^2\over 64\pi^2}\Big(-{1\over 2}+\ln |f''|\Big)$$
 
To study the behavior of symmetry breaking, we must expand around 
$\phi_0$, a 
minimum of $h_0(\phi)$, rather than around $\rho$,
a minimum of the classical potential $U(\phi)$.  Radiative corrections shift 
both the location of the minima and their depth.  The shift in location 
affects the depth at $O(r^3 (\ln |r|)^2)$, and we ignore it.  In our 
nonpolynomial theories, radiative 
corrections always dampen symmetry breaking.  They do not induce it in 
the classically unbroken theories, but sometimes destroy 
it in the classically broken ones.  To analyze this 
we study the difference between $V(0)$ and $V_{min}$
(using $f'_0$ to 
denote $f'(0)$ and $f'$ to denote $f'(\phi_0)$):

\begin{eqnarray}%
\Delta V&=& V(\phi_{min})-V(0)\approx V(\phi_0)-V(0)\\
&=& r\Big[f+ {1\over 32\pi^2}
(f''-f_0'')+(r\ln |r|) {f''^2-f_0''^2\over 
64\pi^2}+r\Big({f''^2\over 64\pi^2}\ln |f''|-
{f_0''^2\over 64\pi^2}\ln |f_0''|\Big)\Big]\\
\end{eqnarray}

\begin{figure}
\centerline{\hbox{
\psfig{file=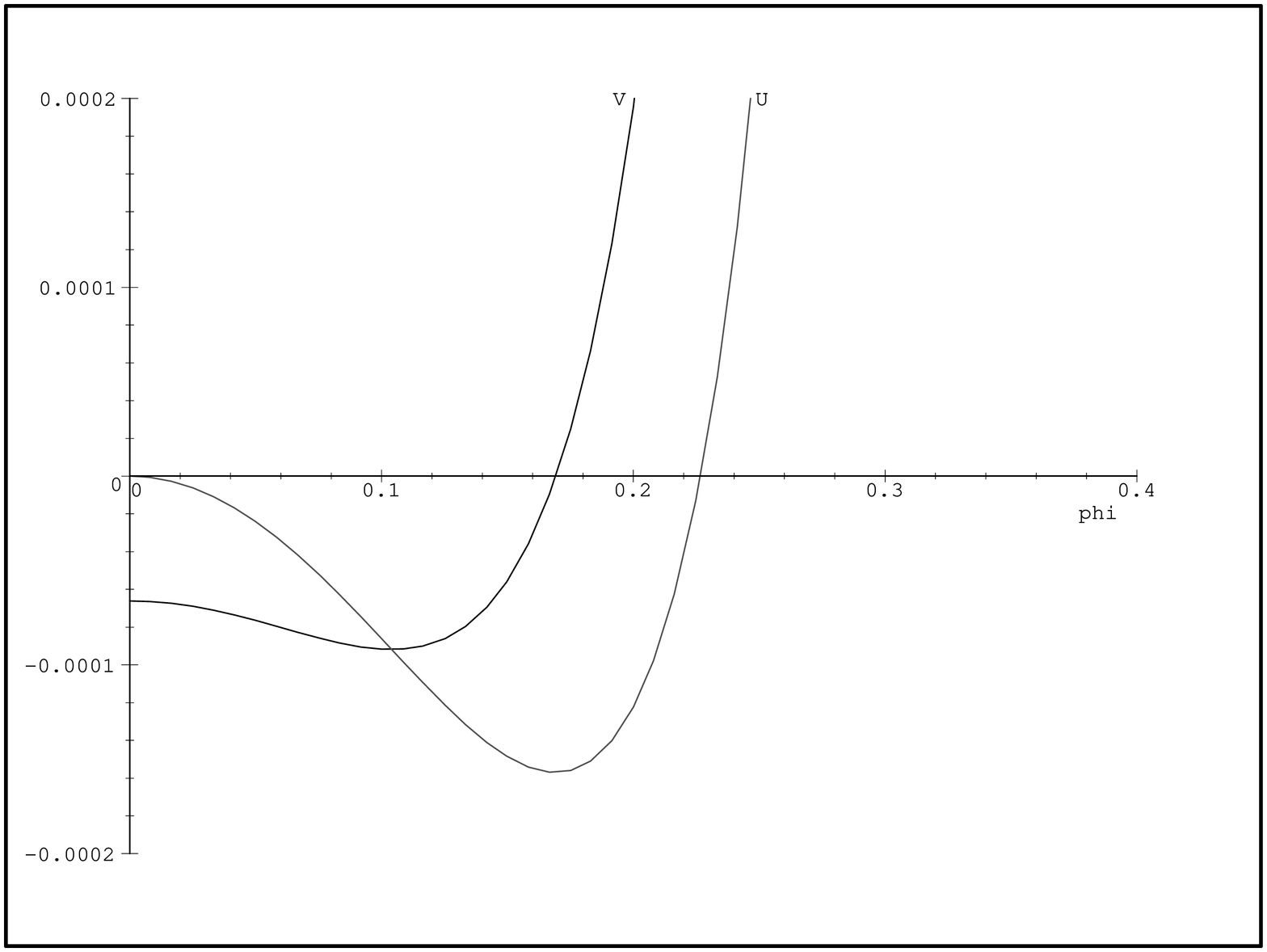,height=1.7truein,width=1.7truein,angle=-90} 
\psfig{file=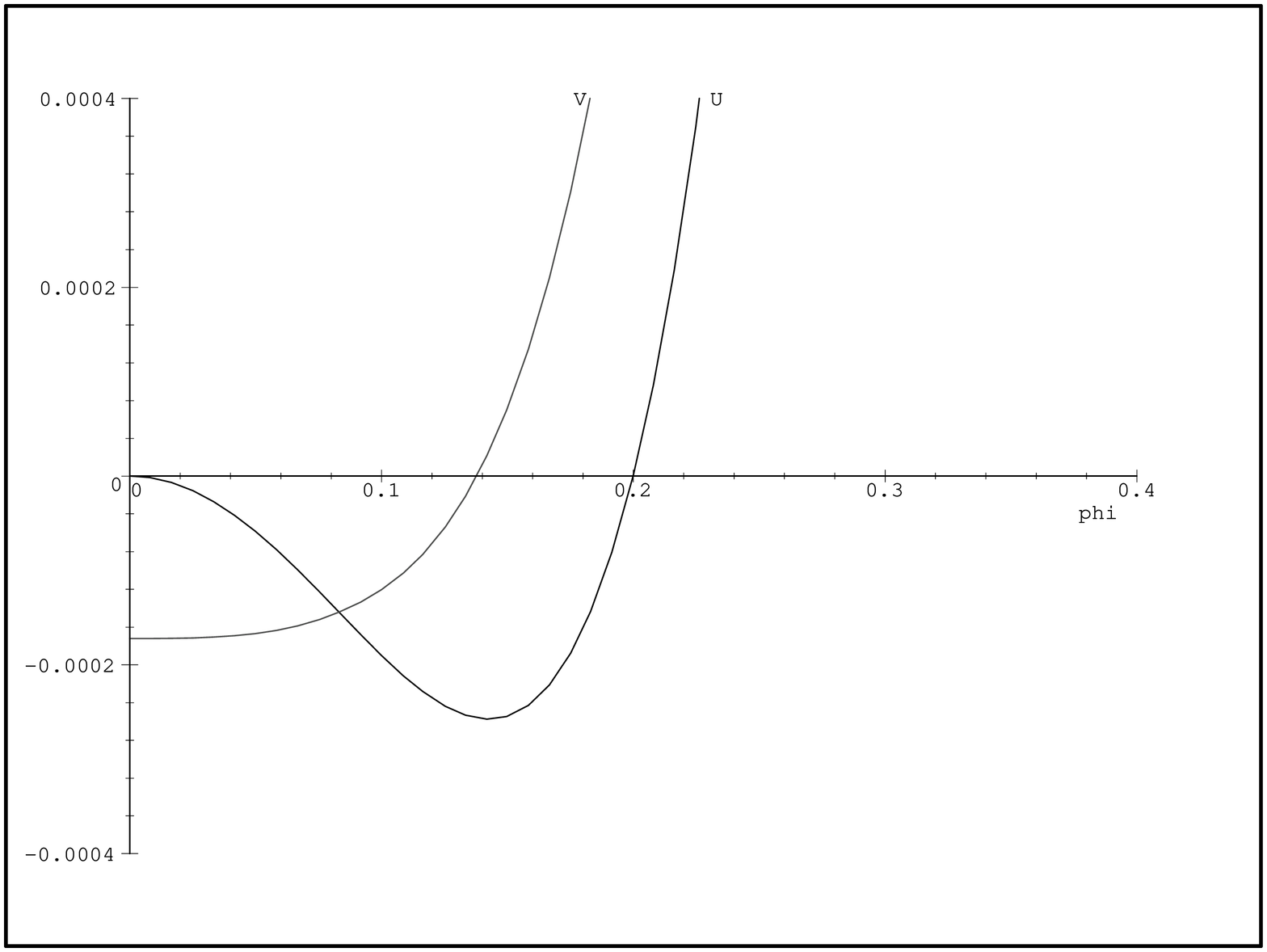,height=1.7truein,width=1.7truein,angle=-90} 
\psfig{file=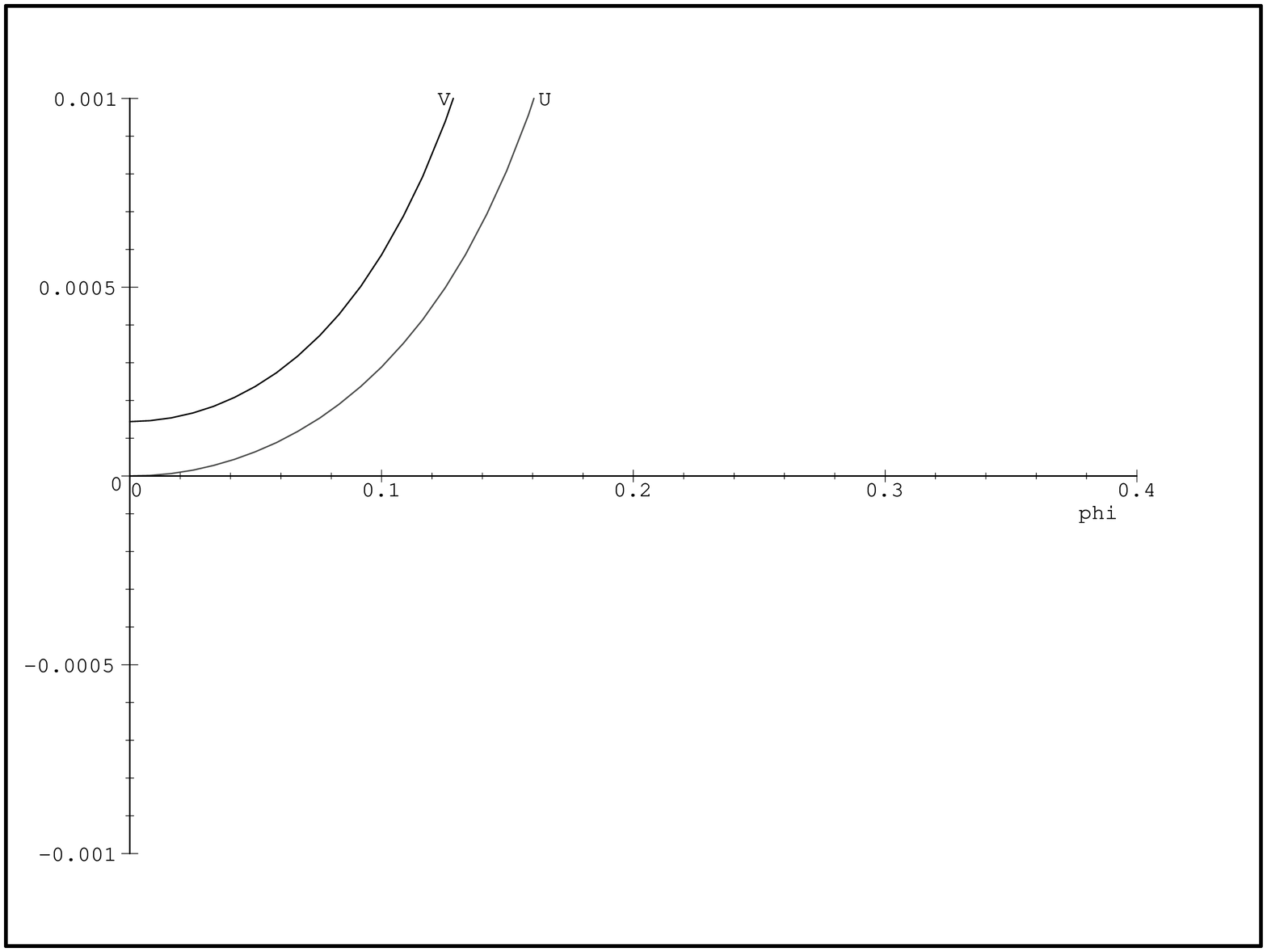,height=1.7truein,width=1.7truein,angle=-90}
}}
\centerline{\hbox{\hskip 0truein (a)\hskip 1.6truein 
(b)\hskip 1.6truein (c)}}
\caption{Comparison of classical potential $U$ and effective potential $V$ 
for  
(a) $r=-0.02$ and $a=-0.5$, (b) $r=-0.05$ and $a=-0.9$,
and (c) $r=0.05$ and $a=0.5$.} 
\label{fig:efftypes}
\end{figure}

\begin{figure}
\vbox{\centerline{\hbox{
\psfig{file=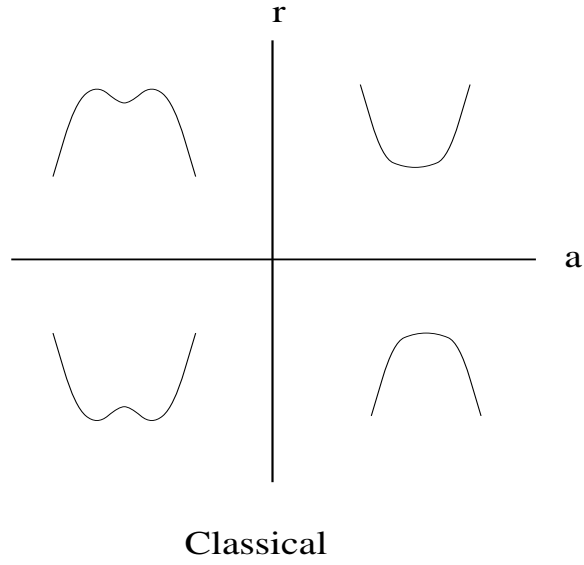,height=3truein,width=3truein,angle=-90}}}
\vskip 0.3truein
\centerline{\hbox{
\psfig{file=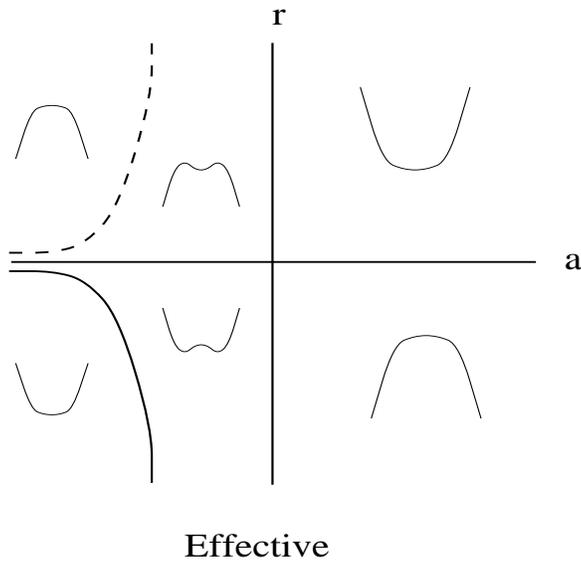,height=3truein,width=3truein,angle=-90}}}}
\caption{Symmetry broken and unbroken regions in classical and $1$-loop
effective theories.} 
\label{fig:phasebound}
\end{figure}

When $\Delta V<0$, there is no symmetry breaking at $1$-loop order.  There
is no closed form expression for $\phi_0$ as a function of $a$ and $r$; so,
analytical calculation proves unenlightening and we must resort to numerical
computation to study $\Delta V$.  
Figure~\ref{fig:efftypes} compares
the classical and effective potentials for three choices of $(r,a)$.  In the 
first case, the effective potential maintains but dampens 
the broken nature of the 
classical potential.  In the second, the classical potential exhibits
symmetry breaking but the effective potential does not.  In the third case, 
the classical potential is unbroken, as is the effective potential.  As is
evident from the second case, radiative corrections can destroy classical
symmetry breaking in our theories.  In fact, in $(r,a)$ parameter space, 
with $r<0$ and $a<0$, there is a phase boundary, plotted in 
figure~\ref{fig:phasebound}, below which radiative effects destroy
symmetry breaking.  To the extent that we can numerically probe, no 
radiatively induced symmetry breaking is present for the classically
unbroken theories $a>0$, $r>0$.  
For $0>a>-0.585$, all theories exhibit symmetry breaking.  
Of course, our results rely on $r$ being small, so we only expect the top
portion of the phase curve to be valid.  Though we cannot probe its entire
structure, we have been alerted to its existence.   

One point requires elucidation.  It may appear strange that symmetry breaking
is most pronounced in the region of $a$-space bounded 
by the free theory ($a=0$)
and not by the $\phi^4$ theory ($a=-1$).  
This is a result of the manner in 
which the eigenpotentials evolve from $\phi^4$ to free theories as $a$ 
progresses from $-1$ to $0$.  For $r<0$, the minima move downward and outward
approaching a negative parabola\footnote{Of course, the free theory is 
really a positive parabola because $r>0$ for an unbroken theory.}
at $a=0$.
This explains why the region near $a=0$ sees the most
pronounced symmetry breaking.      
  
\section{Summary and Conclusions}

In an ordinary polynomial theory, the monomial couplings are of direct 
physical significance, being the lowest order scattering amplitudes.  In a 
nonpolynomial theory, the lowest order 
scattering amplitudes involve an infinite summation of diagrams.  This is 
tantamount to a field renormalization.  
The resulting amplitudes demonstrate the expected asymptotic freedom.  
In symmetry broken theories, however, a divergence remains.  This reflects
an inability to isolate the physics associated with different vacua because
of the steepness of the walls of our exponential potential.  
  
We have computed scattering cross sections to leading order in $r$
and have studied  their high energy scaling.  In contrast to 
pure $\phi^4$ theory, their scaling is of the form 
$\sigma_{2\rightarrow 2}(E)\sim E^{\lambda^2-6\lambda-2}
(\ln E)^{4-\lambda}$.  

Examination of the $1$-loop effective potential has yielded insight into the 
symmetry breaking behavior of our eigenpotentials.  We have found a phase
boundary between symmetry broken and unbroken phases.  If we choose an
eigendirection with 
$-1<a<a_c=-.585$ and scale the energy up (scale $r\rightarrow 0$), we cross
from an unbroken theory to a broken theory.  This is the opposite of the 
usual paradigm in which symmetry is restored at higher energies.  The 
physical implications of this are unclear and merit further
examination. 

One important caveat is in order.  Our original eigenpotential calculation was
valid to linear order in $r$.  Beyond this order, the projection of the flow
in the local, non-derivative subspace is apparently unchanged, 
but nonlocal interactions may arise.  The calculations we have 
performed have been at leading order, not linear order, in $r$.   
However, the lowest order RG corrections that arise from moving 
away from the Gaussian fixed point are $O(r^2)$, which is higher than the 
leading terms in our calculations.  Because the RG does not generate our 
leading terms, it is reasonable to believe our calculations to be valid.

\end{document}